\documentstyle[amsmath,12pt]{article}

\begin{document}

\begin{titlepage}

\begin{center}

{\Large \bf
Relativistic non-instantaneous action-at-a-distance interactions}
\vskip .6in

Domingo J. Louis-Martinez
\vskip .2in

Science One Program and\\ 
Department of Physics and Astronomy,\\ University of British
Columbia\\Vancouver, Canada

\end{center}
 
\vskip 3cm
 
\begin{abstract}
Relativistic action-at-a-distance theories with
interactions that propagate at the speed of light in vacuum are investigated.
We consider the most general action depending on the 
velocities and relative positions of the particles.
The Poincar\'e invariant parameters that label successive events along
the world lines can be identified with the proper times of the particles 
provided that certain conditions are impossed on the interaction
terms in the action.
Further conditions on the interaction terms arise from the requirement that
mass be a scalar. 
A generic class of theories with interactions
that satisfy these conditions is found.
The relativistic equations
of motion for these theories are presented.
We obtain exact circular orbits solutions of the relativistic
one-body problem.
The exact relativistic one-body Hamiltonian is also derived.
The theory has three components: a linearly rising potential, a
Coulomb-like interaction and a dynamical component to the Poincar\'e
invariant  mass.
At the quantum level we obtain the generalized
Klein-Gordon-Fock equation and the Dirac equation.
\end{abstract} 
\end{titlepage}

In the past thirty years or so, a great deal of work has been focused
on the problem of relativistic bound states \cite{anderson} - \cite{deluca},
particularly on the relativistic
equations
for quark-antiquark bound states and the problem of
deriving the meson spectrum\cite{prosperi}, \cite{olsson}.
Within the relativistic
action-at-a-distance formulation of Wheeler and Feynman
\cite{wheeler} - \cite{domingo}
(for electrodynamics) solutions of the two-body Dirac equations
\cite{alstine}, \cite{sazdjian2} have been
found for positronium \cite{alstine}.
The spectrum obtained by this method agrees with
the result of quantum field theory \cite{brezin}, \cite{bethe}
at least up to the $\alpha^{4}$ order.
The approach has also been applied to mesons \cite{crater}.
There is
strong experimental evidence that for large separations the interaction
between quarks can be effectively described by a linearly rising potential
\cite{olsson}.  Several relativistic generalizations of a linearly rising
potential have been studied \cite{katz} - \cite{weiss}.
From quantum chromodynamics, using the Wilson loops approach
\cite{sazdjian}, it has been
established that the quark-antiquark bound states are effectively described
by a static potential, which is a sum of a
linearly rising potential and a Coulomb-like interaction:
$V = \sigma r - \frac{k}{r}$
\cite{allen}.

In this Letter we extend the approach of Wheeler and Feynman
to explore what
types of interparticle interactions are allowed in special relativity.
We assume that the interactions travel at the speed of light in vacuum and
that the theory can be
described by an action principle for which the interaction terms in the
action do not depend on the four-vector accelerations or on higher
derivatives.

We find explicitly the most general theory that satisfies these conditions.
In the static limit we find the theory has three components: a linearly
rising potential, a Coulomb-like interaction and a dynamical component to the
Poincar\'e invariant mass.

We obtain the relativistic equations of motion for $N$ particles
and apply these results to the relativistic one-body problem, for which
we obtain explicitly the Hamiltonian. Quantum mechanical equations for
spinless particles and for spin -$\frac{1}{2}$ particles are presented at
the end of the Letter.
The possibility of considering the effect of a dynamical
component to the quark masses in current phenomenological models is
naturally suggested by the results obtained here.

Let us consider a system of $N$ interacting relativistic
particles. Let $m_{i}$ $(i= 1,2,...,N)$ be the mass
of particle $i$, $c$ is the speed of light and
$\lambda_{i}$ a Poincar\'e invariant parameter labelling the events along
the world line $z^{\mu}_{i}(\lambda_{i})$
of particle $i$ in Minkowski spacetime. We denote
$\dot{z}^{\mu}_{i} = \frac{d z^{\mu}_{i}}{d \lambda_{i}}$.

The metric tensor: $\eta_{\mu\nu}= diag(+1,-1,-1,-1)$. 
We can write the Poincar\'e invariants \cite{schild}, \cite{ramond}:

\begin{equation}
\zeta_{i} = \dot{z}_{i}^{2}
\label{1a}
\end{equation}

\begin{equation}
\xi_{i j}= \left(\dot{z}_{i} \dot{z}_{j}\right) 
\label{1c}
\end{equation}

\begin{equation}
\gamma_{i j}= \left(\dot{z}_{i} (z_{j} - z_{i})\right) 
\label{1d}
\end{equation}

\begin{equation}
\rho_{i j}= \left( z_{i} - z_{j}\right)^{2} 
\label{1f}
\end{equation}

Let us consider the action:

\begin{equation}
S = \sum\limits^{}_{i} m_{i}c\int d\lambda_{i} \zeta_{i} +
\sum\limits^{}_{i}  \sum\limits^{}_{j\neq i} \frac{g_{i} g_{j}}{c}
\int \int d\lambda_{i} d\lambda_{j}
F \left(\xi_{ij}, \gamma_{ij}, \gamma_{j i}, \zeta_{i}, \zeta_{j}\right)
\delta\left(\rho_{ij}\right)
\label{2}
\end{equation}

The Dirac delta function in (\ref{2}) accounts for the interactions
propagating at the speed of light forward and backward in time.

Without loss of generality, we assume the function $F$ to be symmetric:
\begin{equation}
F \left(\xi_{ij}, \gamma_{ij}, \gamma_{j i}, \zeta_{i}, \zeta_{j}\right) =
F \left(\xi_{ij}, \gamma_{ji}, \gamma_{ij}, \zeta_{j}, \zeta_{i}\right)
\label{3}
\end{equation}

The Minkowski equations of motion for N interacting relativistic particles
can be derived from the action (\ref{2}) using the variational
principle. We find \footnote{Equations (\ref{4} - \ref{6c}) are derived
from (\ref{2}) by varying
$z_{i\mu}$ in the action and integrating by parts, taking into
account that:

$\frac{d}{d \lambda_{i}} \left(\delta(\rho_{ij})\right) =
\frac{\left(\frac{d \rho_{ij}}{d \lambda_{i}}\right)}
{\left(\frac{d \rho_{ij}}{d \lambda_{j}}\right)} 
\frac{d}{d \lambda_{j}} \left(\delta(\rho_{ij})\right) =
\frac{\gamma_{ij}}{\gamma_{ji}}
\frac{d}{d \lambda_{j}} \left(\delta(\rho_{ij})\right)$}:

\begin{equation}
m_{i}\ddot{z}^{\mu}_{i} = K^{\mu}_{i}
\label{4}
\end{equation}

\noindent where,

\begin{eqnarray}
K^{\mu}_{i} & = &
\frac{g_{i}}{c^2} \sum\limits^{}_{j\neq i}  g_{j}
\int d\lambda_{j} \delta\left(\rho_{ij}\right)
\left(A^{\mu}_{i j} + B^{\mu \nu}_{i j} \ddot{z}_{i \nu} +
C^{\mu \nu}_{i j} \ddot{z}_{j \nu} \right)
\label{5}
\end{eqnarray}

\begin{eqnarray}
A^{\mu}_{i j} & = & \frac{\partial F}{\partial z_{i \mu}} -
\frac{\partial^{2} F}{\partial z^{\eta}_{i}\partial \dot{z}_{i \mu}}
\dot{z}^{\eta}_{i} +
\frac{\zeta_{j}}{\gamma_{ji}^{2}} \left( (z^{\mu}_{i} - z^{\mu}_{j}) F +
\gamma_{ij} \frac{\partial F}{\partial\dot{z}_{i \mu}}\right) \nonumber\\
& & + \frac{1}{\gamma_{ji}} \left( - \dot{z}^{\mu}_{j} F +
(z^{\mu}_{i} - z^{\mu}_{j}) \frac{\partial F}{\partial z^{\eta}_{j}}
\dot{z}^{\eta}_{j} + \xi_{ij} \frac{\partial F}{\partial \dot{z}_{i \mu}}
+ \gamma_{ij} \frac{\partial^{2} F}{\partial z^{\eta}_{j}
\partial \dot{z}_{i \mu}} \dot{z}^{\eta}_{j}\right)
\label{6a}
\end{eqnarray}

\begin{equation}
B^{\mu \nu}_{i j} =
- \frac{\partial^{2} F}{\partial \dot{z}_{i \mu} \partial \dot{z}_{i \nu}} 
\label{6b}
\end{equation}

\begin{equation}
C^{\mu \nu}_{i j} =
\frac{(z^{\mu}_{i} -z^{\mu}_{j})}{\gamma_{ji}}
\left(\frac{\partial F}{\partial \dot{z}_{j \nu}}
- \frac{(z^{\nu}_{i} -z^{\nu}_{j})}{\gamma_{ji}} F \right)
+ \frac{\gamma_{ij}}{\gamma_{ji}} \left(
\frac{\partial^{2} F}{\partial \dot{z}_{i \mu} \partial \dot{z}_{j \nu}}
- \frac{(z^{\nu}_{i} -z^{\nu}_{j})}{\gamma_{ji}}
\frac{\partial F}{\partial \dot{z}_{i \mu}}
\right)
\label{6c}
\end{equation}

In order to identify $\lambda_{i}$ with $s_{i} = c\tau_{i}$, where $\tau_{i}$
is the particle's proper time in flat spacetime, one needs to impose the
well known conditions:
\begin{equation}
\left(K_{i} \dot{z}_{i}\right) = 0
\label{7}
\end{equation}

The conditions (\ref{7}) guarantee that, for all solutions of the equations
of motion, $\zeta_{i}$  $(i = 1,2,...,N)$ are constants (which by simple scaling can be
made equal to $1$):

\begin{equation}
d\lambda_{i}^{2} = \eta_{\mu\nu} d z^{\mu}_{i} d z^{\nu}_{i}
\label{8}
\end{equation}

Taking into account Eqs (\ref{5}) and (\ref{6a}, \ref{6b}, \ref{6c}) we
can see that Eqs (\ref{7}) lead to the following conditions on $F$:

\begin{equation}
\gamma_{ji} \dot{z}_{i\eta} \frac{\partial \tilde{F}}{\partial z_{i\eta}} -
\gamma_{ij} \dot{z}_{j\eta} \frac{\partial\tilde{F}}{\partial z_{j\eta}} -
\left(\xi_{ij} + \zeta_{j} \frac{\gamma_{ij}}{\gamma_{ji}}\right) \tilde{F}
= 0
\label{9a}
\end{equation}

\begin{equation}
\frac{\partial \tilde{F}}{\partial \dot{z}_{i\nu}} = 0
\label{9b}
\end{equation}

\begin{equation}
(z^{\nu}_{i} - z^{\nu}_{j}) \tilde{F} -
\gamma_{ji} \frac{\partial \tilde{F}}{\partial \dot{z}_{j\nu}} = 0
\label{9c}
\end{equation}

\noindent where,

\begin{equation}
\tilde{F} =
F - \dot{z}_{i\mu} \frac{\partial F}{\partial \dot{z}_{i\mu}}
\label{10}
\end{equation}

$F$ is a function of the invariants $\xi_{ij}$, $\gamma_{ij}$, $\gamma_{ji}$,
$\zeta_{i}$ and $\zeta_{j}$.
From (\ref{9b}) it follows that $\tilde{F}$ does not depend
on $\xi_{ij}$, $\gamma_{ij}$ and $\zeta_{i}$. Therefore, we can rewrite
the conditions (\ref{9a}) and (\ref{9c}) on
$\tilde{F}\left(\gamma_{ji}, \zeta_{j}\right)$ as follows:

\begin{equation}
\left(\xi_{ij} + \zeta_{j} \frac{\gamma_{ij}}{\gamma_{ji}}\right)
\left(\gamma_{ji}\frac{\partial \tilde{F}}{\partial \gamma_{ji}} -
\tilde{F} \right) = 0
\label{11}
\end{equation}

\begin{equation}
(z^{\nu}_{i} - z^{\nu}_{j})
\left(\gamma_{ji}\frac{\partial\tilde{F}}{\partial \gamma_{ji}} -
\tilde{F}\right) +
2 \dot{z}^{\nu}_{j}\gamma_{ji} \frac{\partial \tilde{F}}{\partial \zeta_{j}}
= 0
\label{12}
\end{equation}

From (\ref{11}) and (\ref{12}) it follows that
$\tilde{F}$ must satisfy the partial
differential equations:

\begin{equation}
\gamma_{ji}\frac{\partial\tilde{F}}{\partial\gamma_{ji}} - \tilde{F} = 0
\label{13}
\end{equation}

\begin{equation}
\frac{\partial \tilde{F}}{\partial \zeta_{j}} = 0
\label{14}
\end{equation}

From (\ref{13}) and (\ref{14}) it follows that $\tilde{F}$ must be of
the form: $\tilde{F} = \kappa \gamma_{ji}$, where $\kappa$ is a constant.
Therefore, from this result and (\ref{10}), it follows that $F$ must obey the
following two conditions:

\begin{equation}
F - \xi_{ij} \frac{\partial F}{\partial \xi_{ij}} -
\gamma_{ij} \frac{\partial F}{\partial \gamma_{ij}} -
2 \zeta_{i} \frac{\partial F}{\partial \zeta_{i}} = \kappa \gamma_{ji}
\label{15a}
\end{equation}

\begin{equation}
F - \xi_{ij} \frac{\partial F}{\partial \xi_{ij}} -
\gamma_{ji} \frac{\partial F}{\partial \gamma_{ji}} -
2 \zeta_{j} \frac{\partial F}{\partial \zeta_{j}} = \kappa \gamma_{ij}
\label{15b}
\end{equation}

The most general solution to the system of partial differential
equations (\ref{15a}) and (\ref{15b}) can be written in the form:

\begin{equation}
F = \kappa \left(\gamma_{ij} + \gamma_{ji}\right) +
f\left(\xi_{ij}, \zeta_{ij}, \eta_{ij}, \eta_{ji}\right)
\label{16}
\end{equation}

\noindent where,

\begin{equation}
\zeta_{ij} = \sqrt{\zeta_{i} \zeta_{j}}
\label{17a}
\end{equation}

\begin{equation}
\eta_{ij} = \frac{\gamma_{ij}}{\sqrt{\zeta_{i}}}
\label{17b}
\end{equation}

\begin{equation}
\eta_{ji} = \frac{\gamma_{ji}}{\sqrt{\zeta_{j}}}
\label{17c}
\end{equation}

\noindent and the function $f$ is homogeneous of degree one in the first two
arguments and symmetric:

\begin{equation}
f\left(t\xi_{ij}, t\zeta_{ij}, \eta_{ij}, \eta_{ji}\right) =
t f\left(\xi_{ij}, \zeta_{ij}, \eta_{ij}, \eta_{ji}\right)
\label{18}
\end{equation}

\begin{equation}
f\left(\xi_{ij}, \zeta_{ij}, \eta_{ji}, \eta_{ij}\right) =
f\left(\xi_{ij}, \zeta_{ij}, \eta_{ij}, \eta_{ji}\right)
\label{19}
\end{equation}

We have considered the most general action depending on the velocities and
relative positions for $N$ relativistic interacting particles (\ref{2}).
We have found that, in order to identify the Poincar\'e invariant parameters
$\lambda_{i}$ with $s_{i} = c \tau_{i}$,
the interation terms in the action must be of the form (\ref{16}).

The equations of motion (\ref{4}) can be rewritten as follows:

\begin{equation}
\bar{M}^{\mu\nu}_{i} \ddot{z}_{i\nu} = \bar{K}^{\mu}_{i}
\label{20}
\end{equation}

\noindent where,

\begin{equation}
\bar{M}^{\mu\nu}_{i} = m_{i} \eta^{\mu\nu} - 
\frac{g_{i}}{c^2} \sum\limits^{}_{j\neq i}  g_{j}
\int d s_{j} \delta\left(\rho_{ij}\right)
B^{\mu \nu}_{i j} 
\label{21}
\end{equation}

\begin{equation}
\bar{K}^{\mu}_{i} =
\frac{g_{i}}{c^2} \sum\limits^{}_{j\neq i}  g_{j}
\int d s_{j} \delta\left(\rho_{ij}\right)
\left(A^{\mu}_{i j} +
C^{\mu \nu}_{i j} \ddot{z}_{j \nu} \right)
\label{22}
\end{equation}

Notice that the four-vector force $\bar{K}_{i}$ does not depend on
$\ddot{z}_{i}$.

In (\ref{20}) $\bar{M}_{i}$ can be interpreted as the particle's mass,
a tensor under Poincar\'e's transformations.

If we assume mass to be a scalar quantity, which we denote as $\bar{m}_{i}$
$(i = 1, 2, ..., N)$, then the tensor $\bar{M}_{i}$ should be of the form:

\begin{equation}
\bar{M}^{\mu\nu}_{i} = \bar{m}_{i} \eta^{\mu\nu} +
\bar{n}^{\mu}_{i} \dot{z}^{\nu}_{i}
\label{24}
\end{equation}

From the expressions (\ref{21}), (\ref{6b}) and (\ref{16})
we find that the requirement
of a Poincar\'e
invariant mass (\ref{24}) translates into the following new conditions on
the interaction function $f$:

\begin{equation}
\frac{\partial^{2} f}{\partial \eta_{ij}^{2}} = 0
\label{26a}
\end{equation}

\begin{equation}
\frac{\partial^{2} f}{\partial \eta_{ji}^{2}} = 0
\label{26b}
\end{equation}

\begin{equation}
\frac{\partial^{2} f}{\partial \xi_{ij}^{2}} = 0
\label{26c}
\end{equation}

\begin{equation}
\frac{\partial^{2} f}{\partial \xi_{ij}\partial \eta_{ij}} = 0
\label{26d}
\end{equation}

\begin{equation}
\frac{\partial^{2} f}{\partial \xi_{ij}\partial \eta_{ji}} = 0
\label{26e}
\end{equation}

\begin{equation}
\frac{\partial}{\partial \xi_{ij}}
\left(\zeta_{ij}\frac{\partial f}{\partial \zeta_{ij}} -
\eta_{ij}\frac{\partial f}{\partial \eta_{ij}}\right) = 0
\label{26f}
\end{equation}

\begin{equation}
\frac{\partial}{\partial \xi_{ij}}
\left(\zeta_{ij}\frac{\partial f}{\partial \zeta_{ij}} -
\eta_{ji}\frac{\partial f}{\partial \eta_{ji}}\right) = 0
\label{26g}
\end{equation}

\begin{equation}
\frac{\partial}{\partial \eta_{ij}}
\left(\zeta_{ij}\frac{\partial f}{\partial \zeta_{ij}} -
\eta_{ij}\frac{\partial f}{\partial \eta_{ij}}\right) = 0
\label{26h}
\end{equation}

\begin{equation}
\frac{\partial}{\partial \eta_{ji}}
\left(\zeta_{ij}\frac{\partial f}{\partial \zeta_{ij}} -
\eta_{ji}\frac{\partial f}{\partial \eta_{ji}}\right) = 0
\label{26i}
\end{equation}

The most general solution to the system of differential equations
(\ref{26a} - \ref{26i}) that also satisfies the conditions
(\ref{18}) and (\ref{19}) can be written in the following form:

\begin{equation}
f = \alpha \zeta_{ij} \eta_{ij} \eta_{ji} + \beta \xi_{ij}
+ \gamma \zeta_{ij}
+ \delta \zeta_{ij} \left(\eta_{ij} + \eta_{ji}\right)
\label{27}
\end{equation}

\noindent where $\alpha$, $\beta$, $\gamma$ and $\delta$ are constants.

Substituting (\ref{16}) and (\ref{27}) into the expression (\ref{2}) for the
action functional we obtain:

\begin{eqnarray}
S & = & \sum\limits^{}_{i} m_{i}c\int d\lambda_{i} \zeta_{i} + 
\sum\limits^{}_{i}  \sum\limits^{}_{j\neq i} \frac{g_{i} g_{j}}{c}
\int \int d\lambda_{i} d\lambda_{j}  \delta\left(\rho_{ij}\right)
\left(\left(\kappa + \delta \zeta_{j}^{\frac{1}{2}} \right) \gamma_{ij}
+ \left(\kappa + \delta \zeta_{i}^{\frac{1}{2}}\right)\gamma_{ji}\right)
\nonumber\\
& & + \sum\limits^{}_{i}  \sum\limits^{}_{j\neq i} \frac{g_{i} g_{j}}{c}
\int \int d\lambda_{i} d\lambda_{j}  \delta\left(\rho_{ij}\right)
\left( \alpha \gamma_{ij} \gamma_{ji} + \beta \xi_{ij}
+ \gamma \zeta_{i}^{\frac{1}{2}} \zeta_{j}^{\frac{1}{2}}\right)
\label{28}
\end{eqnarray}

Notice that the second term in (\ref{28}) vanishes identically, since:

\begin{equation}
\int d \lambda_{j} \delta \left(\rho_{ij}\right) \gamma_{ji} =
\frac{1}{2} \sum\limits^{}_{s = -, +} \int d \lambda_{j}
\delta\left(\lambda_{j} - \lambda^{(i, s)}_{j}\right)
\frac{\gamma_{ji}}{|\gamma_{ji}|} = 0
\label{29}
\end{equation}

In (\ref{29}), $\lambda^{(i, s)}_{j}$ are the two roots
of the equation:

\begin{equation}
\left(z_{i}(\lambda_{i}) - z_{j}(\lambda^{(i,s)}_{j})\right)^{2} = 0
\label{30}
\end{equation}

Therefore, we have found that the most general theory with an action
functional of the form (\ref{2})
for which:

1. the Poincar\'e invariant parameters $\lambda_{i} = s_{i} = c \tau_{i}$,

2. all the masses are scalars under Poincar\'e transformations,

\noindent is given by the action:

\begin{equation}
S = \sum\limits^{}_{i} m_{i}c\int d\lambda_{i} \zeta_{i} + 
+ \sum\limits^{}_{i}  \sum\limits^{}_{j\neq i} \frac{g_{i} g_{j}}{c}
\int \int d\lambda_{i} d\lambda_{j}  \delta\left(\rho_{ij}\right)
\left( \alpha \gamma_{ij} \gamma_{ji} + \beta \xi_{ij}
+ \gamma \zeta_{i}^{\frac{1}{2}} \zeta_{j}^{\frac{1}{2}}\right)
\label{31}
\end{equation}

From the action (\ref{31}), we obtain the equations of motion for
a system of $N$ interacting relativistic particles:

\begin{equation}
\bar{m}_{i} \ddot{z}^{\mu}_{i} = \bar{K}^{\mu}_{i}
\label{32}
\end{equation}

The Poincare invariant masses are equal to:

\begin{equation}
\bar{m}_{i} = m_{i} + \gamma \frac{g_{i}}{c^2} \sum\limits^{}_{j\neq i} g_{j}
\int d s_{j} \delta\left((z_{i} - z_{j})^2\right)
\label{32a}
\end{equation}

The four-vector forces $\bar{K}^{\mu}_{i},  (i=1,2,...,N)$ can be expressed
in the compact form:

\begin{equation}
\bar{K}^{\mu}_{i} = g_{i} \left(F^{\mu\nu}_{i} \dot{z}_{i\nu}
+ \Gamma^{\mu}_{i \alpha\beta} \dot{z}^{\alpha}_{i}\dot{z}^{\beta}_{i}\right)
\label{33}
\end{equation}

\noindent where $F^{\mu\nu}_{i}$ is an antisymmetric tensor
($F^{\nu\mu}_{i} = - F^{\mu\nu}_{i}$) given by the
following formula:

\begin{eqnarray}
F^{\mu\nu}_{i} & = & \sum\limits^{}_{j\neq i} \frac{g_{j}}{c^{2}} \int d s_{j}
\frac{\delta\left(\rho_{ij}\right)}
{\gamma_{ji}^{2}} 
\left[ \alpha \gamma_{ji}^{2}
\left[(z^{\mu}_{i} - z^{\mu}_{j})\dot{z}^{\nu}_{j} -
\dot{z}^{\mu}_{j} (z^{\nu}_{i} - z^{\nu}_{j})\right]  \right. \nonumber\\
& & + \beta
\left[\left((z^{\mu}_{i} - z^{\mu}_{j})\dot{z}^{\nu}_{j} -
\dot{z}^{\mu}_{j} (z^{\nu}_{i} - z^{\nu}_{j})\right)
\left(1 - \sigma_{ji}\right) +
\left. \left((z^{\mu}_{i} - z^{\mu}_{j})\ddot{z}^{\nu}_{j} -
\ddot{z}^{\mu}_{j} (z^{\nu}_{i} - z^{\nu}_{j})\right)
\gamma_{ji}\right]\right] \nonumber\\
\label{33a}
\end{eqnarray}

\noindent and $\Gamma^{\mu}_{i\alpha\beta}$ is a symmetric tensor
in flat space-time
($\Gamma^{\mu}_{i\alpha\beta} = \Gamma^{\mu}_{i\beta\alpha}$) given by
the expression:

\begin{eqnarray}
\Gamma^{\mu}_{i\alpha\beta} & = &
\gamma \sum\limits^{}_{j\neq i} \frac{g_{j}}{c^{2}} \int d s_{j}
\frac{\delta\left(\rho_{ij}^{2}\right)}
{\gamma_{ji}^{2}} \nonumber\\
& & \left[\left((z^{\mu}_{i} - z^{\mu}_{j})\eta_{\alpha\beta} -
\frac{1}{2}\left(\delta^{\mu}_{\alpha}(z_{i\beta} - z_{j\beta})
+ \delta^{\mu}_{\beta}(z_{i\alpha} - z_{j\alpha})\right)\right)
\left(1 - \sigma_{ji}\right)\right. \nonumber\\
& & -\left. \left(\dot{z}^{\mu}_{j}\eta_{\alpha\beta} -
\frac{1}{2}(\delta^{\mu}_{\alpha}\dot{z}_{j\beta}
+ \delta^{\mu}_{\beta}\dot{z}_{j\alpha})\right)\gamma_{ji}
\right]
\label{33b}
\end{eqnarray}

In (\ref{33a}, \ref{33b}):

\begin{equation}
\sigma_{ji} = \left(\ddot{z}_{j} (z_{i} - z_{j})\right)
\label{34}
\end{equation}

The relativistic equations of motion 
(\ref{32} - \ref{33b})
admit exact solutions for any number of particles.
The details of the calculation will be given in a future
publication.

In the remaining part of this Letter we will consider the relativistic
one-body problem. 

Consider a particle of mass $m_{1} = m$ and charge
$g_{1} = g$ interacting with a heavy particle (of mass $m_{2} \gg m_{1}$ and
charge $g_{2} = \tilde{g}$) at rest
($\vec{v}_{2} = 0$). We obtain the equations of motion:

\begin{eqnarray}
& & \left(m + \frac{\gamma g \tilde{g}}{c^{2} r}\right)
\left(\vec{a}
+ \frac{(\vec{v} \vec{a})}{c^{2} \left(1 - \frac{v^{2}}{c^{2}}\right)}
\vec{v}\right) = \nonumber\\
& & \frac{g \tilde{g}}{r} \left[
\left(1 - \frac{v^{2}}{c^{2}}\right)^{\frac{1}{2}}
\left(\alpha + \frac{\beta}{r^{2}}\right) \vec{r}
+ \frac{\gamma}{r^{2}} \left(
\left(1 - \frac{v^{2}}{c^{2}}\right) \vec{r} +
\frac{(\vec{r} \vec{v})}{c^{2}} \vec{v}\right)\right]
\label{A1}
\end{eqnarray}

These equations follow from (\ref{32} - \ref{33b}).

The one-body equations of motion (\ref{A1}) can also be derived from the
Lagrangian:

\begin{equation}
L = - m c^{2} \left(1 + \frac{\gamma g \tilde{g}}{m c^{2} r}\right)
\left(1 - \frac{v^{2}}{c^{2}}\right)^{\frac{1}{2}}
+ g \tilde{g} \left(\alpha r - \frac{\beta}{r}\right)
\label{A2}
\end{equation}

The exact relativistic one-body Hamiltonian is given by the expression:

\begin{equation}
H = m c^{2} \left(\left(1 + \frac{\gamma g \tilde{g}}{m c^{2} r}\right)^{2}
+ \frac{p^{2}}{m^{2} c^{2}}\right)^{\frac{1}{2}}
- g \tilde{g} \left(\alpha r - \frac{\beta}{r}\right)
\label{A3}
\end{equation}

The conserved energy $E$ and angular momentum $\vec{L}$ can be found to be:

\begin{equation}
E = \frac{\left(m c^{2} + \frac{\gamma g \tilde{g}}{r}\right)}
{\left(1 - \frac{v^{2}}{c^{2}}\right)^{\frac{1}{2}}}
- g \tilde{g} \left(\alpha r - \frac{\beta}{r}\right)
\label{A4}
\end{equation}

\begin{equation}
\vec{L} = \frac{\left(m + \frac{\gamma g \tilde{g}}{c^{2} r}\right)}
{\left(1 - \frac{v^{2}}{c^{2}}\right)^{\frac{1}{2}}}
\vec{r} \times \vec{v}
\label{A5}
\end{equation}

The motion of the relativistic particle is planar, with polar coordinates
$(r, \phi)$ obeying the equations:

\begin{equation}
\frac{d \phi}{d t} = \frac{c^{2} L}
{r^{2} \left(E + g \tilde{g} \left(\alpha r - \frac{\beta}{r}\right)\right)}
\label{A6}
\end{equation}

\begin{equation}
\frac{1}{c^{2}} \left(\frac{d r}{d t}\right)^{2} =
1 - \frac{\left(\left(m + \frac{\gamma g \tilde{g}}{c^{2} r}\right)^{2} c^{4}
+ \frac{L^{2} c^{2}}{r^{2}}\right)}
{\left(E + g \tilde{g} \left(\alpha r - \frac{\beta}{r}\right)\right)^{2}}
\label{A7}
\end{equation}

The equations of motion admit circular orbits solutions with the angular
frequency $\omega$ and the radius $r$ obeying the generalized Kepler
relation:

\begin{equation}
\omega^{2} = - \frac{g \tilde{g}}{m r^{3}}
\left[\gamma + \left(\alpha r^{2} + \beta\right) \left(
\left(
1 + \frac{\gamma g \tilde{g}}{m c^{2} r}
+ \frac{g^{2} \tilde{g}^{2}(\alpha r^{2} + \beta)^{2}}{4 m^{2} c^{4} r^{2}}
\right)^{\frac{1}{2}}
+ \frac{g \tilde{g}(\alpha r^{2} + \beta)}{2 m c^{2} r} 
\right)
\right]
\label{A8}
\end{equation}

For circular orbits the energy is given by the expression:

\begin{equation}
E = m c^{2} \left(
1 + \frac{\gamma g \tilde{g}}{m c^{2} r}
+ \frac{g^{2} \tilde{g}^{2}(\alpha r^{2} + \beta)^{2}}{4 m^{2} c^{4} r^{2}}
\right)^{\frac{1}{2}}
+ \frac{g \tilde{g}}{2} \left( - 3\alpha r + \frac{\beta}{r}\right)
\label{A9}
\end{equation}

\noindent and the angular momentum is:
 
\begin{equation}
L = \left(1 + \frac{\gamma g \tilde{g}}{m c^{2} r}\right)
\frac{\left[
- g \tilde{g} m r
\left[\gamma +
\left(\alpha r^{2} + \beta\right)
\left(
\left(1 + \frac{\gamma g \tilde{g}}{m c^{2} r}
+ \frac{g^{2} \tilde{g}^{2}(\alpha r^{2} + \beta)^{2}}
{4 m^{2} c^{4} r^{2}}
\right)^{\frac{1}{2}}
+ \frac{g \tilde{g}(\alpha r^{2} + \beta)}{2 m c^{2} r}
\right)
\right]
\right]^{\frac{1}{2}}}
{\left[
\left(1 + \frac{\gamma g \tilde{g}}{m c^{2} r}
+ \frac{g^{2} \tilde{g}^{2}(\alpha r^{2} + \beta)^{2}}{4 m^{2} c^{4} r^{2}}
\right)^{\frac{1}{2}}
+ \frac{g \tilde{g}(\alpha r^{2} + \beta)}{2 m c^{2} r}
\right]}
\label{A10}
\end{equation}

At the quantum level for spinless particles we obtain
the relativistic Klein-Gordon-Fock equation as follows:

\begin{equation}
\left(i \hbar \frac{\partial}{\partial t}
+ g \tilde{g} \left(\alpha r - \frac{\beta}{r}\right)\right)^{2}
\Psi\left(t, \vec{r}\right) =
m^{2} c^{4} \left(1 + \frac{\gamma g \tilde{g}}{m c^{2} r}\right)^{2}
\Psi\left(t, \vec{r}\right) - \hbar^{2} c^{2} \nabla^{2}
\Psi\left(t, \vec{r}\right)
\label{A13}
\end{equation}

For stationary states,
$\Psi (t, \vec{r}) = e^{- \frac{i E t}{\hbar}} \psi(\vec{r})$,
in spherical coordinates we can separate the radial and angular parts of
the wave function in the standard way:

\begin{equation}
\psi(\vec{r}) = R(r) Y_{l m}\left(\theta, \phi\right)
\label{A15}
\end{equation}

\noindent where $Y_{l m}\left(\theta, \phi\right)$ are the spherical
harmonics. For the radial part we obtain the equation:

\begin{equation}
\left(- \frac{1}{r^{2}} \frac{d}{d r}\left(r^{2} \frac{d}{d r}\right)
+ \frac{l (l + 1)}{r^{2}}\right) R =
\frac{1}{\hbar^{2} c^{2}} \left[
\left(E + g \tilde{g} \left(\alpha r - \frac{\beta}{r}\right)\right)^{2} -
m^{2} c^{4} \left(1 + \frac{\gamma g \tilde{g}}{m c^{2} r}\right)^{2}\right]
R
\label{A16}
\end{equation}

For spin- $\frac{1}{2}$ particles we can write the Dirac equation as follows:

\begin{equation}
i \hbar \frac{\partial}{\partial t} \Psi = \hat{H} \Psi
\label{A17}
\end{equation}

\noindent where,

\begin{equation}
\hat{H} = i \hbar c \vec{\alpha}_{o} \vec{\nabla}
- \beta_{o} m \left(1 + \frac{\gamma g \tilde{g}}{m c^{2} r}\right)
- g \tilde{g} \left(\alpha r - \frac{\beta}{r}\right)
\label{A18}
\end{equation}

One can easily check that the total angular momentum
$\vec{L} + \frac{1}{2} \hbar \vec{\sigma}_{o}$ commutes with the
Hamiltonian operator (\ref{A18}).

The matrices $\vec{\alpha}_{o}$, $\beta_{o}$
and $\vec{\sigma}_{o}$
may be written
in the standard way in terms of the Pauli matrices $\vec{\sigma}$ and the
unit matrix $I$.

\end{document}